\documentclass{article}
 \usepackage[dblblindworkshop, preprint]{neurips_2025}
\workshoptitle{Differentiable Systems and Scientific Machine Learning}

\usepackage[utf8]{inputenc} 
\usepackage[T1]{fontenc}    
\usepackage[colorlinks=true, linkcolor=., citecolor=black, urlcolor=blue]{hyperref}       
\usepackage{url}            
\usepackage{booktabs}       
\usepackage{nicefrac}       
\usepackage{microtype}      
\usepackage{xcolor}         
\usepackage{amsmath}
\usepackage{amssymb}
\usepackage{mathtools, bm}
\usepackage{multicol}
\usepackage{multirow}
\usepackage{graphicx}
\usepackage{siunitx}
\usepackage{enumitem}
\usepackage{import}
\usepackage{siunitx}
\usepackage{listings}
\usepackage[linesnumbered,lined,boxed,commentsnumbered]{algorithm2e}
\usepackage{pgfplots}
\usepgfplotslibrary{units, groupplots}
\pgfplotsset{unit markings=parenthesis, compat=1.18}

\definecolor{codegreen}{rgb}{0,0.6,0}
\definecolor{codegray}{rgb}{0.5,0.5,0.5}
\definecolor{codepurple}{rgb}{0.58,0,0.82}
\definecolor{backcolour}{rgb}{0.95,0.95,0.92}

\lstdefinestyle{mystyle}{
    backgroundcolor=\color{backcolour},   
    commentstyle=\color{codegreen},
    keywordstyle=\color{blue},
    numberstyle=\tiny\color{codegray},
    stringstyle=\color{codepurple},
    basicstyle=\ttfamily\footnotesize,
    breakatwhitespace=false,         
    breaklines=true,                 
    captionpos=b,                    
    keepspaces=true,                 
    numbersep=5pt,                  
    showspaces=false,                
    showstringspaces=false,
    showtabs=false,                  
    tabsize=2
}

\RestyleAlgo{ruled}

\newcommand{\RR}{I\!\!R}

\newcommand{\LL}{\mathcal{L}}

\newcommand{\bx}{\mathbf{x}}

\newcommand{\bv}{\mathbf{v}}
\newcommand{\bz}{\mathbf{z}}

\newcommand{\bB}{\mathbf{B}}
\newcommand{\bC}{\mathbf{C}}
\newcommand{\bA}{\mathbf{A}}

\newcommand{\bI}{\mathbf{I}}

\title{Accelerating Automatic Differentiation of \\
Direct Form Digital Filters}

\author{
  Chin-Yun Yu \qquad  Gy\"orgy Fazekas \\
  Centre for Digital Music\\
  Queen Mary University of London\\
  Mile End Road, London E1 4NS, UK \\
  \texttt{\{\href{mailto:chin-yun.yu@qmul.ac.uk}{chin-yun.yu}, 
    \href{mailto:george.fazekas@qmul.ac.uk}{george.fazekas}\}@qmul.ac.uk} \\
}

\begin{document}

\maketitle

\begin{abstract}
    We introduce a general formulation for automatic differentiation through direct form filters, yielding a closed-form backpropagation that includes initial condition gradients.
    The result is a single expression that can represent both the filter and its gradients computation while supporting parallelism.
    C++/CUDA implementations in PyTorch achieve at least 1000x speedup over naive Python implementations and consistently run fastest on the GPU.
    For the low-order filters commonly used in practice, exact time-domain filtering with analytical gradients outperforms the frequency-domain method in terms of speed.
\end{abstract}

\section{Introduction}
\label{sec:intro}

Direct form (DF) digital filters with coefficients derived directly from their rational transfer functions are widely used in signal processing.
With the advent of deep learning, there is increasing interest in integrating these filters into neural networks as an inductive bias and jointly optimising them using gradient-based methods.
Differentiable implementations have been used for equaliser matching~\cite{nercessian_neural_nodate}, filter design\cite{colonel_direct_2021}, virtual analogue modelling~\cite{kuznetsov_differentiable_nodate, nercessian_lightweight_2021}, or bigger end-to-end systems like differentiable music tracks mixing~\cite{steinmetz_style_2022,lee2025reverse,diffvox}.
All-pole filters, a special case of DF filters, are fundamental building blocks for voice modelling~\cite{markel_linear_1976} and their differentiable version is used in neural vocoders and voice synthesis~\cite{valin_lpcnet_2019, subramani_end--end_2022, schulze2023unsupervised, yu_singing_2023, ycy2024golfv1}.
For a comprehensive review of differentiable filters, refer to Table 2 in \cite{hayes2024review}.
Outside audio applications, differentiable DF filters have also been used in system identification~\cite{forgione_dynonet_2021} and state-space sequence models~\cite{pmlr-v235-parnichkun24a}.

Efficiency is a significant issue when training filters in popular automatic differentiation (AD) frameworks such as PyTorch~\cite{paszke_pytorch_2019}.
At the time of writing, PyTorch native operators only support filters with no recursions, such as convolution.
Evaluating recursive DF filters using the available differentiable operators is slow due to the added up function call overheads in Python~\cite{nercessian_lightweight_2021,yu_singing_2023,ycy2024golfv1}.
This is not the case for other AD frameworks that have recursion operators, such as \texttt{jax.lax.scan} in JAX~\cite{jax2018github} and \texttt{tf.scan} in TensorFlow~\cite{tensorflow2015-whitepaper}.
To address this limitation, a common workaround is to approximate the filter using frequency sampling (FS)~\cite{hayes2024review}, thereby leveraging the fast Fourier transform (FFT) available in AD frameworks.
Nevertheless, sampling creates time-domain aliasing~\cite{MDFTWEB07} where its infinite impulse response (IIR) is folded.
To reduce this mismatch error, one can increase the FFT resolution~\cite{lee2022differentiable} or sample the transfer function outside the unit circle~\cite{dal2025flamo}.
Parnichkun et al.~\cite{pmlr-v235-parnichkun24a} show that the tail of the IIR can be removed entirely in FS by subtracting it with a decayed and delayed version of itself.
In addition, some works perform filtering directly using FS~\cite{nercessian_lightweight_2021, steinmetz_style_2022, juvela2019gelp,oh_excitglow_2020, wright2022grey,carson2023differentiable}, which results in circular convolution with the folded response, introducing more errors.

In this paper, we propose registering filters as low-level operators in PyTorch and deriving analytical gradients for backpropagation, thereby bypassing the overheads of AD frameworks.
This approach has been used for specific filter types~\cite{diffvox, yu_singing_2023,ycy2024golfv1,forgione_dynonet_2021}, but we generalise it to all possible direct form filters.
This general form includes the transposed direct form (TDF), which has better numerical properties than DF and is the standard implementation in scientific computing libraries, thus better aligning with the needs of the open science community.
It also allows exploration of acceleration techniques on parallel hardware, such as GPUs, which are not discussed in prior work.
We stick with the time-domain implementation since filters used in practice are usually low-order, and the FS method is not necessarily faster in this case.
We also show the analytical form of the gradients for the initial conditions, which can be beneficial when optimising on short sequences~\cite{FORGIONE202169}.
Our implementation is published in the open-source package \texttt{philtorch}.\footnote{\href{https://github.com/yoyolicoris/philtorch}{github.com/yoyolicoris/philtorch}}

\section{Background}
\label{sec:background}

An $M^{\mathrm{th}}$-order time-invariant linear system has the following transfer function in the $z$-domain:
\begin{equation}
    \label{eq:rtf}
    H(z) = \frac{b_0 + b_1 z^{-1} + b_2 z^{-2} + \ldots + b_M z^{-M}}{1 + a_1 z^{-1} + a_2 z^{-2} + \ldots + a_M z^{-M}}
\end{equation}
where $b_i, a_i$ are the feed-forward and feedback coefficients, respectively.
To simplify the notation, we assume the numerator and denominator have the same order (padding if necessary).
Directly converting the filtered signal $Y(z)=H(z)X(z)$ into difference equations results in the so-called \emph{Direct-Form}\footnote{For simplicity, we assume type-II (transposed) direct forms throughout this paper.} implementation:
\begin{align}
    v(n) & = x(n) - a_1 v(n-1) - a_2 v(n-2) - \ldots - a_M v(n-M)     \label{eq:df_in} \\
    y(n) & = b_0 v(n) + b_1 v(n-1) + \ldots + b_M v(n-M)   \label{eq:df_out}
\end{align}
where $x(n)$ and $y(n)$ are the input and output signals, respectively.
However, it is more common to see the \emph{Transposed-Direct-Form} implementation in practice, such as in SciPy~\cite{2020SciPy-NMeth} and MATLAB, because the interleaved $b_i$ coefficients provide compensating attenuation, making them more numerically robust~\cite{NUMROBUSTTDFII07}.
To see how DF and TDF are related, let us utilise the following state-space filter:
\begin{align}
    \bv(n+1) & = \bA \bv(n) + \bB x(n) \label{eq:ss_recursion} \\
    y(n)     & = \bC^\top \bv(n) + D x(n) \label{eq:ss_output}
\end{align}
where $\bv(n) \in \RR^M$ is the state vector, $\bA \in \RR^{M \times M}$ is the transition matrix, $\bB, \bC \in \RR^M$ are the input and output matrices, and $D \in \RR$ is the direct path coefficient.
The initial state $\bv(0)$ gives an entry point to initiate the recursion.
When $\bA$ is the companion matrix of the polynomial in the denominator of Eq.~\eqref{eq:rtf}, $\bC = [b_1 - a_1 b_0, b_2 - a_2 b_0, \ldots, b_M - a_M b_0]^\top$, $\bB = [1, 0, \ldots, 0]^\top$, and $D = b_0$, Eq.~\eqref{eq:ss_recursion} and Eq.~\eqref{eq:ss_output} are equivalent to Eq.~\eqref{eq:df_in} and Eq.~\eqref{eq:df_out}.
TDF traverses the signal flow of DF in reverse, which does not alter the transfer function~\cite{TRANSPOSEDFORMS07}.
Representing TDF in state-space form is simply replacing $\bA$ with its transpose and swapping $\bB$ and $\bC$~\cite{TRANSPOSESSM07}.

Efficient AD through DF was developed separately by Forgione~\cite{forgione_dynonet_2021} and Yu~\cite{yu_singing_2023,ycy2024golfv1}.
They treat Eq.~\eqref{eq:df_out} and Eq.~\eqref{eq:df_in} as separate filters and derive analytical gradients for the latter.
This idea was later extended to parameter-varying all-pole filters and improved in~\cite{ycy2024golf, ycy2024diffapf}, where the backpropagation is further simplified.
The latter version is implemented in TorchAudio~\cite{torchaudio}.
However, in TDF, the feed-forward and feedback coefficients are interleaved in the difference equations~\cite{TRANSPOSEDFORMS07}, making it non-trivial to apply the same technique.
In this work, we start from the state-space form, as it represents the coefficients in two matrices, making it easier to derive gradients.

\section{Methodology}
\label{sec:method}

\subsection{Backpropagation through state-space filters}
\label{ssec:backprop}
Since we are implementing the filter without AD for optimal speed, we need to compute the filter's gradient analytically.
We outline the results in this section.
For detailed derivations, please refer to Appendix~\ref{appendix:derivation}.
Given a scalar value $\LL = f(y(0), y(1), \dots)$ we wish to minimise, computed via a differentiable function $f$, we aim to calculate the gradients of the parameters we are interested in with respect to it.
The parameters can be the filter variables or other parameters that the variables depend on.
In reverse-mode AD, the gradients with respect to the output $\frac{\partial \LL}{\partial y(n)}$ are provided.
Our task is to compute $\frac{\partial \LL}{\partial x(n)}$, $\frac{\partial \LL}{\partial \bv(0)}$, $\frac{\partial \LL}{\partial \bA}$, $\frac{\partial \LL}{\partial \bB}$, $\frac{\partial \LL}{\partial \bC}$, and $\frac{\partial \LL}{\partial D}$.
The instantaneous gradients $\frac{\partial \LL}{\partial \bv(n)}$ and $\frac{\partial \LL}{\partial \bC}$, $\frac{\partial \LL}{\partial D}$ are trivial to compute:
\begin{align}
    \label{eq:dLdCD}
    \frac{\partial \LL}{\partial \bv(n)} = \frac{\partial \LL}{\partial y(n)} \bC^\top, \quad
    \frac{\partial \LL}{\partial \bC}    = \sum_{n \geq 0} \frac{\partial \LL}{\partial y(n)} \bv(n)^\top, \quad
    \frac{\partial \LL}{\partial D} = \sum_{n \geq 0} \frac{\partial \LL}{\partial y(n)} x(n).
\end{align}
For $\frac{\partial \LL}{\partial x(n)}$, let us denote $\bz(n) = \bB x(n)$ so Eq.~\eqref{eq:ss_recursion} becomes $\bv(n+1) = \bA \bv(n) + \bz(n)$.
Given the instantaneous gradients $\frac{\partial \LL}{\partial \bv(n)}$, the gradients with respect to $\bz(n)$ can be computed via \emph{backpropagation-through-time} (BPTT) algorithm~\cite{58337} and have recursive definition:
\begin{equation}
    \label{eq:dLdz}
    \begin{split}
        \frac{\partial \LL}{\partial \bz(n)} & = \left(\bA^\top {\frac{\partial \LL}{\partial \bz(n+1)}^\top} + {\frac{\partial \LL}{\partial \bv(n+1)}^\top}\right)^\top
        = \left(\bA^\top {\frac{\partial \LL}{\partial \bz(n+1)}^\top} + \bC \frac{\partial \LL}{\partial y(n+1)}\right)^\top.
    \end{split}
\end{equation}
The rest of the gradients can be computed as:
\begin{gather}
    \frac{\partial \LL}{\partial x(n)} = \bB {\frac{\partial \LL}{\partial \bz(n)}^\top}  + D \frac{\partial \LL}{\partial y(n)}, \label{eq:dLdx}\\
    \frac{\partial \LL}{\partial \bv(0)} = \frac{\partial \LL}{\partial \bz(-1)}, \quad
    \frac{\partial \LL}{\partial \bB}  = \sum_{n \geq 0} \frac{\partial \LL}{\partial \bz(n)} x(n), \quad
    \frac{\partial \LL}{\partial \bA}  = \sum_{n \geq 0}\bv(n)  \frac{\partial \LL}{\partial \bz(n)}. \label{eq:dLdvBA}
\end{gather}
Notice that Eq.~\eqref{eq:dLdz} and Eq.~\eqref{eq:dLdx} combined form a TDF filter as we have seen in Section~\ref{sec:background}, but running backwards in time since $n$ is decreasing in Eq.~\eqref{eq:dLdz}.
Similarly, if we derive the gradients for the TDF filter, we obtain a DF filter for backpropagation.
We verified that our analytical gradients match the numerical gradients.

\subsection{Implementation considerations}
\label{ssec:proposed}
As we have mentioned in Section~\ref{sec:intro}, recursions like Eq.~\eqref{eq:ss_recursion} and Eq.~\eqref{eq:dLdz} are better made as native operators in AD frameworks.
Given the relation between TDF and DF, we should implement both and register the backpropagation of one using the other to save efforts.
Alternatively, we can implement the state-space filter and register its backpropagation using itself with reparametrised arguments and reverse-time filtering.
The state-space form also enables parallelisation across the time dimension, which is not possible with scalar difference equations.

The recursion Eq.~\eqref{eq:ss_recursion} can be accelerated using the \emph{associative scan} algorithm~\cite{BlellochTR90}, which is implemented in JAX and TensorFlow but still work-in-progress for PyTorch~\cite{wu2025control}.
The trick is to express the recursion as associative operations that can be computed in parallel.
Let us use $\oplus$ to denote a binary operator that merges two tuples $(\bA, \bz) \oplus (\bA', \bz') \mapsto (\bA'\bA, \bA' \bz + \bz')$.
Then, we can express Eq.~\eqref{eq:ss_recursion} as:
\begin{equation}
    \left(\mathbf{0}, \bv(n)\right) = (\mathbf{0}, \bv(0)) \oplus (\bA, \bz(0)) \oplus (\bA, \bz(1)) \oplus \ldots \oplus (\bA, \bz(n-1)).
\end{equation}
Using a parallelised associative scan, $\bv(n)$ only needs $O(\frac{N}{p} + \log(p))$ time to be computed for all $n$ instead of $O(N)$ time, where $N$ is the sequence length and $p$ is the number of parallel workers.
However, the extra multiplications in $\oplus$ can introduce significant overhead when $M$ is large.
Yu et al.~\cite{diffvox} proposed decomposing $\bA$ into diagonal and invertible matrices, reducing matrix multiplications to element-wise multiplications.
Similar ideas have been widely used to accelerate linear RNNs~\cite{martin2018parallelizing} and state-space sequence models such as Mamba~\cite{gu2024mamba}, LRU~\cite{orvieto2023lru}, and S5~\cite{smith2023simplified}.
It is only applicable when $\bA$ is diagonalisable (i.e., no repeated poles).

\section{Experiments}
\label{sec:benchmarks}

To examine the proposed method, we benchmarked the recursion Eq.~\eqref{eq:ss_recursion} with $M = 2$ in PyTorch.\footnote{Specifically, the recursion of $\bv(n+1) = \bA \bv(n) + \bz(n)$ as AD through $\bB \bx(n)$ can be done via PyTorch.}
We choose second-order because it is the minimum order for a real filter to have complex poles, and higher-order filters are often composed of many second-order sections to reduce numerical errors.
The signal length $N$ varies from $2^{14}$ to $2^{20}$, corresponding to 1 to 60 seconds of audio at \SI{16}{\kHz} sample rate, which covers most of the range of audio processing tasks.
We implement the recursion in C++ and CUDA as custom extensions (\textbf{EXT}) and register the backpropagation based on the equations in Section~\ref{ssec:backprop}.
In the C++ implementation, recursion is implemented sequentially, whereas for the CUDA kernel, we utilise the associative scan from the CUDA Core Compute Libraries.\footnote{\href{https://nvidia.github.io/cccl/}{nvidia.github.io/cccl/}}
The diagonalised version of EXT is denoted as \textbf{Diag-EXT}.
The baselines we compare with are all implemented using PyTorch's Python API.
The first one, \textbf{Naive}, is just a simple for-loop.
The second baseline \textbf{Unrolled} implements the parallel scan using differentiable matrix multiplications~\cite{ycy2025unrollssm}.
The third one \textbf{RTF} evaluates the truncated impulse response and performs convolution in the frequency domain using FFT~\cite{pmlr-v235-parnichkun24a}.

\begin{figure}[!t]
    \centering
    \begin{tikzpicture}[scale=1]
    \begin{groupplot}[group style={
                    group size= 2 by 1,
                    horizontal sep=0.05\textwidth,
                    xlabels at=edge bottom,
                    y descriptions at=edge left,
                },
            width=0.4\textwidth,
            height=0.4\textwidth,
            xlabel={\small $N$},
            ylabel={\small Runtime},
            y unit= \mu s,
            tick label style={font=\footnotesize},
            ymajorgrids=true,
            grid style=dashed,
            every axis plot post/.append style={thick},
            ymode=log,
            xmode=log,
            log basis x=2,
            ymin=100, ymax=4000000,
            xmin=10000, xmax=1200000,
        ]
        \nextgroupplot[
            title={\normalsize CPU},
        ]

        \addplot[smooth, color=blue, mark=x, mark size=2pt] coordinates {
                (16384, 1127.6)
                (65536, 1295)
                (262144, 2420.1)
                (1048576, 7840.1)
            };
        \addplot[smooth, color=magenta, mark=square*, mark size=2pt] coordinates {
                (16384, 185.7)
                (65536, 618.8)
                (262144, 2350.1)
                (1048576, 9752.5)
            };
        \addplot[smooth, color=black, mark=triangle*, mark size=2pt] coordinates {
                (16384, 308)
                (65536, 623.3)
                (262144, 1939.5)
                (1048576, 7821.6)
            };
        \addplot[smooth, color=brown, mark=star, mark size=2pt] coordinates {
                (16384, 4993.3)
                (65536, 24599.5)
                (262144, 105172.7)
                (1048576, 551687.1)
            };

        \addplot[smooth, color=red, mark=diamond*, mark size=2pt] coordinates {
                (16384, 56939.9)
                (65536, 240635.8)
                (262144, 983126.1)
                (1048576, 3919639.9)
            };

        \addplot[smooth, color=blue, mark=x, mark size=2pt, dashed] coordinates {
                (16384, 2660.4)
                (65536, 3089.3)
                (262144, 5500)
                (1048576, 22500)
            };
        \addplot[smooth, color=magenta, mark=square*, mark size=2pt, dashed] coordinates {
                (16384, 385.8)
                (65536, 924.8)
                (262144, 3500)
                (1048576, 27600)
            };
        \addplot[smooth, color=black, mark=triangle*, mark size=2pt, dashed] coordinates {
                (16384, 727.9)
                (65536, 1347.8)
                (262144, 8600)
                (1048576, 42600)
            };
        \addplot[smooth, color=brown, mark=star, mark size=2pt, dashed] coordinates {
                (16384, 5355.3)
                (65536, 33013.9)
                (262144, 134556.8)
                (1048576, 647790.5)
            };
        \addplot[smooth, color=red, mark=diamond*, mark size=2pt, dashed] coordinates {
                (16384, 870527.3)
                (65536, 13719452.1)
            };

        \nextgroupplot[title={\normalsize GPU},
            legend style={legend columns=1, font=\footnotesize,
                    at={(1.05,0.5)}, anchor=west},
        ]
        \addplot[smooth, color=red, mark=diamond*, mark size=2pt] coordinates {
                (16384, 318085.4)
                (65536, 1288979.4)
                (262144, 5201269.3)
                (1048576, 20966551.5)
            };
        \addlegendentry{Naive}
        \addplot[smooth, color=blue, mark=x, mark size=2pt] coordinates {
                (16384, 1964)
                (65536, 2503.9)
                (262144, 2617.4)
                (1048576, 3103.2)
            };
        \addlegendentry{Unrolled}
        \addplot[smooth, color=brown, mark=star, mark size=2pt] coordinates {
                (16384, 535.5)
                (65536, 572.4)
                (262144, 879.6)
                (1048576, 3812.3)
            };
        \addlegendentry{RTF}
        \addplot[smooth, color=black, mark=triangle*, mark size=2pt] coordinates {
                (16384, 460.5)
                (65536, 440.9)
                (262144, 452.7)
                (1048576, 670)
            };
        \addlegendentry{Diag-EXT}
        \addplot[smooth, color=magenta, mark=square*, mark size=2pt] coordinates {
                (16384, 135)
                (65536, 141)
                (262144, 155.3)
                (1048576, 213.6)
            };
        \addlegendentry{EXT}


        \addplot[smooth, color=red, mark=diamond*, mark size=2pt, dashed] coordinates {
                (16384, 808895.8)
                (65536, 3297630.6)
            };
        \addlegendentry{Naive (backprop)}
        \addplot[smooth, color=blue, mark=x, mark size=2pt, dashed] coordinates {
                (16384, 6192.7)
                (65536, 7921)
                (262144, 8305.9)
                (1048576, 9634.1)
            };
        \addlegendentry{Unrolled (backprop)}
        \addplot[smooth, color=brown, mark=star, mark size=2pt, dashed] coordinates {
                (16384, 1325.7)
                (65536, 1445.9)
                (262144, 1529.9)
                (1048576, 3907.7)
            };
        \addlegendentry{RTF (backprop)}
        \addplot[smooth, color=black, mark=triangle*, mark size=2pt, dashed] coordinates {
                (16384, 1106.4)
                (65536, 1690.8)
                (262144, 4018)
                (1048576, 27381.8)
            };
        \addlegendentry{Diag-EXT (backprop)}
        \addplot[smooth, color=magenta, mark=square*, mark size=2pt, dashed] coordinates {
                (16384, 451.5)
                (65536, 463)
                (262144, 709.6)
                (1048576, 2394.9)
            };
        \addlegendentry{EXT (backprop)}


    \end{groupplot}
\end{tikzpicture}
    \caption{Runtime of the recursion Eq.~\eqref{eq:ss_recursion} and its backpropagation with various signal lengths $N$ and implementations.
    }
    \label{fig:bench}
\end{figure}
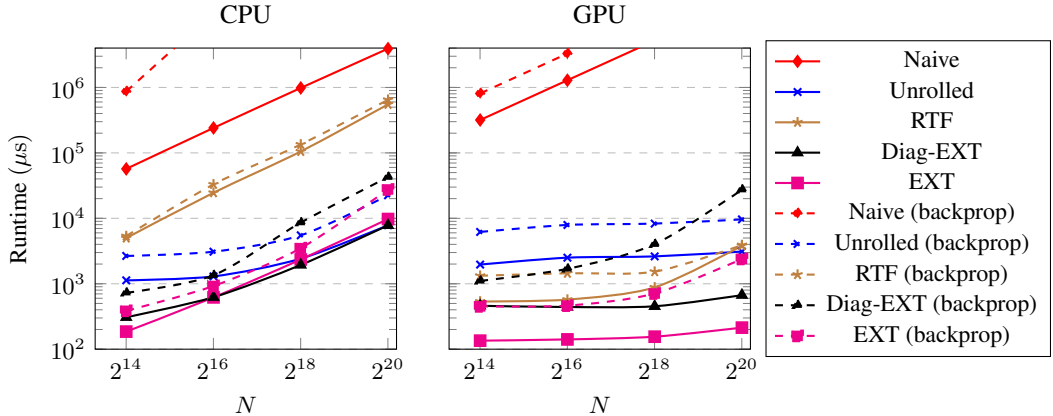

The benchmarks were run with PyTorch 2.8 and CUDA 12.9, using single precision, on a Ubuntu 25.04 machine equipped with an Intel i7-7700K CPU and an NVIDIA RTX 5060 Ti GPU.
We limited the CPU benchmarks to a single thread to see the speed-up from having custom extensions without parallel acceleration.
The results are shown in Figure~\ref{fig:bench}.
Our methods are the most efficient on both platforms, and EXT consistently spends the least time computing gradients in most configurations.
The naive implementation is remarkably slower than others and quickly becomes impractical when $N$ increases.
On the CPU, the unrolled method has higher overhead for smaller $N$ but becomes comparable to the other two methods for large $N$.
The RTF method is only slightly faster than Naive.
On the GPU, RTF becomes comparable to Diag-EXT in smaller values of $N$ but grows faster as $N$ increases and exceeds Unrolled at $N=2^{20}$.
The gap between Diag-EXT and EXT on the GPU is likely due to the extra eigen-space projection in Diag-EXT, and we expect the gap to disappear as $M$ increases.

\section{Conclusion}
\label{sec:conclusion}

We presented the general form of automatic differentiation via (transposed) direct form filters in the state-space domain.
The derivations yield a closed-form backward pass that is itself a state-space (DF/TDF) recursion, and include gradients with respect to initial conditions.
This view enables a single low-level operator whose backpropagation is implemented by the same kernel.
Furthermore, the state-space form naturally supports parallelisation.
Our custom PyTorch extensions that implement the proposed recursion substantially outperform both the naive and frequency-sampling baselines.
These results reinforce the conclusion that exact time-domain filtering with analytical gradients is preferable for low-order filters.
Additionally, we notice that the presented derivation can be easily extended to parameter-varying cases and applies to a broader class of linear state-space models.
Moreover, a forward-mode differentiation can be derived similarly, which is helpful for second-order optimisation methods.
How the closed-form solution affects the numerical accuracy of gradients is also an interesting direction to explore.
We left these extensions to future work.

\begin{ack}
    Chin-Yun Yu is a research student at the UKRI Centre for Doctoral Training in Artificial Intelligence and Music, supported jointly by UK Research and Innovation (grant number EP/S022694/1) and Queen Mary University of London.
    Gy\"orgy Fazekas's research on Knowledge-driven Deep Learning for Music Informatics was supported by the Leverhulme Trust and the Royal Academy of Engineering under the RAEng / Leverhulme Trust Research Fellowships scheme.
\end{ack}

\bibliographystyle{unsrt}
\bibliography{ref}

\appendix

\section{Derivation details}
\label{appendix:derivation}
\subsection{$\frac{\partial \LL}{\partial \bz(n)}$}

This section provides the derivation details for Eq.~\eqref{eq:dLdz} where $\bz(n) = \bB x(n)$.
Let us unroll Eq.~\eqref{eq:ss_recursion} so only $\bv(0)$ and $\bz(n)$ are on the right-hand side for any non-negative integer $n$:
\begin{equation}
    \label{eq:unrolled_ssm}
    \bv(n+1) = \bA^{n+1} \bv(0) + \sum_{m=0}^{n} \bA^{n-m} \bz(m).
\end{equation}
Let $\bv(0) = \sum_{m=-\infty}^{-1} \bA^{-m-1} \bz(m)$, which is always valid, as a solution of $\bz(m)$ for $m < 0$ always exists that produces the same $\bv(0)$ with any $\bA$.
Then, we can eliminate $\bv(0)$ so Eq.~\eqref{eq:unrolled_ssm} is simplified as:
\begin{equation}
    \label{eq:unrolled_ssm_infinite}
    \bv(n) = \bz(n-1) + \sum_{m=-\infty}^{n-2} \bA^{n-m-1} \bz(m).
\end{equation}
This form does not need to account for boundary conditions, which makes the following derivation cleaner.
We can interpret Eq.~\eqref{eq:unrolled_ssm_infinite} as the response of the system to an infinite-length input sequence.

Then, it is easy to see that the Jacobian of $\bv(n)$ with respect to $\bz(n)$ is:
\begin{equation}
    \frac{\partial \bv(n)}{\partial \bz(m)} = \begin{cases}
        \bA^{n-m-1} & m < n - 1 \\
        \bI         & m = n - 1 \\
        \mathbf{0}  & m \geq n
    \end{cases},
\end{equation}
Given the instantaneous gradients $\frac{\partial \LL}{\partial \bv(n)}$ computed in Eq.~\eqref{eq:dLdCD}, we can compute $\frac{\partial \LL}{\partial \bz(n)}$ using the chain rule:
\begin{equation}
    \label{eq:dLdz_full}
    \begin{split}
        \frac{\partial \LL}{\partial \bz(n)}
         & = \sum_{m=-\infty}^\infty \frac{\partial \LL}{\partial \bv(m)} \frac{\partial \bv(m)}{\partial \bz(n)}                                                              \\
         & = \frac{\partial \LL}{\partial \bv(n+1)} + \sum_{m = n + 2}^\infty \frac{\partial \LL}{\partial \bv(m)} \bA^{m-n-1}                                                 \\
         & = \left({\frac{\partial \LL}{\partial \bv(n+1)}}^\top + \sum_{m = n + 2}^\infty \left(\bA^\top\right)^{m-n-1} \frac{\partial \LL}{\partial \bv(m)}^\top\right)^\top \\
         & = \left(\bA^\top {\frac{\partial \LL}{\partial \bz(n+1)}^\top} + {\frac{\partial \LL}{\partial \bv(n+1)}^\top}\right)^\top.
    \end{split}
\end{equation}
The last step is obtained by noticing that the term in the parentheses can be converted from Eq.~\eqref{eq:unrolled_ssm_infinite}, with $\bA$ replaced by $\bA^\top$, $\bz(m)$ replaced by $\frac{\partial \LL}{\partial \bv(m)}^\top$, and $m$ replaced by $2n - m$.
The $m \mapsto 2n - m$ reparametrisation implies reverse-time filtering.
In practice, we cannot evaluate infinite sequences but up to a finite length $N$ (from $y(0)$ to $y(N-1)$).
Thus, $\frac{\partial \LL}{\partial \bz(n)}$ is zeros for $n \geq N-1$ since $\bz(n)$ are not evaluated beyond $n = N - 2$.
We can set $\frac{\partial \LL}{\partial \bz(N-1)} = \mathbf{0}$ to initiate the evaluation of Eq.~\eqref{eq:dLdz_full} from $n = N-1$ to $n = 0$.

\subsection{$\frac{\partial \LL}{\partial \bA}$}
To derive $\frac{\partial \LL}{\partial \bA}$, let us utilise the property of the Kronecker product that $\bA \bv(n) = \text{vec}(\bA\bv(n)) = (\bv(n)^\top \otimes \bI) \text{vec}(\bA)$, where $\text{vec}(\cdot)$ denotes the vectorisation operator that stacks the columns of a matrix into a vector.
The Jacobian of $\bv(n)$ with respect to $\text{vec}(\bA)$ can be expressed as:
\begin{equation}
    \begin{split}
        \frac{\partial \bv(n)}{\partial \text{vec}(\bA)}
         & = \frac{\partial \bv(n)}{\partial \bv(n-1)} \frac{\partial \bv(n-1)}{\partial \text{vec}(\bA)} + \bv(n-1)^\top \otimes \bI \\
         & = \bA \frac{\partial \bv(n-1)}{\partial \text{vec}(\bA)} + \bv(n-1)^\top \otimes \bI                                       \\
         & = \sum_{m=0}^{n-1} \bA^{n-m-1} \left(\bv(m)^\top \otimes \bI\right).
    \end{split}
\end{equation}
Then, applying the chain rule, we have:
\begin{equation}
    \begin{split}
        \frac{\partial \LL}{\partial \text{vec}(\bA)}
         & = \sum_{m = 1}^\infty \frac{\partial \LL}{\partial \bv(m)} \frac{\partial \bv(m)}{\partial \text{vec}(\bA)}
        \\
         & = \sum_{m = 1}^\infty \frac{\partial \LL}{\partial \bv(m)} \sum_{n=0}^{m-1} \bA^{m-n-1} \left(\bv(n)^\top \otimes \bI\right)
        \\
         & =  \left(\sum_{m = 1}^\infty \left(\sum_{n=0}^{m-1} \left(\bv(n) \otimes \bI\right) \left(\bA^\top\right)^{m-n-1}\right) \frac{\partial \LL}{\partial \bv(m)}^\top\right)^\top
        \\
         & =  \left(\sum_{n = 0}^\infty \left(\bv(n) \otimes \bI\right) \sum_{m=n+1}^{\infty} \left(\bA^\top\right)^{m-n-1} \frac{\partial \LL}{\partial \bv(m)}^\top\right)^\top
        \\
         & =  \left(\sum_{n = 0}^\infty \left(\bv(n) \otimes \bI\right) \left(
            \frac{\partial \LL}{\partial \bv(n+1)}^\top + \sum_{m=n+2}^{\infty} \left(\bA^\top\right)^{m-n-1} \frac{\partial \LL}{\partial \bv(m)}^\top\right)\right)^\top
        \\
         & =  \left(\sum_{n = 0}^\infty \left(\bv(n) \otimes \bI\right) {\frac{\partial \LL}{\partial \bz(n)}^\top}\right)^\top
        \\
         & = \left(\sum_{n = 0}^\infty \text{vec}\left(\frac{\partial \LL}{\partial \bz(n)}^\top \bv(n)^\top\right)\right)^\top
        \\
         & =  \text{vec}\left(\sum_{n = 0}^\infty \frac{\partial \LL}{\partial \bz(n)}^\top \bv(n)^\top\right)^\top.
    \end{split}
\end{equation}
The conversion from the fifth to the sixth step is based on Eq.~\eqref{eq:dLdz_full}.
The second last step utilises the same property we mentioned before that $\text{vec}(\mathbf{U} \mathbf{W}) = (\mathbf{W}^\top \otimes \bI) \text{vec}(\mathbf{U})$.
After removing the vectorisation operator on both sides, we obtain $\frac{\partial \LL}{\partial \bA}$ shown in Eq.~\eqref{eq:dLdvBA}.

\subsection{$\frac{\partial \LL}{\partial \bv(0)}$}
From Eq.~\eqref{eq:unrolled_ssm}, it is easy to see that the Jacobian of $\bv(n)$ with respect to $\bv(0)$ is:
\begin{equation}
    \frac{\partial \bv(n)}{\partial \bv(0)} = \bA^n.
\end{equation}
Apply the chain rule, we have:
\begin{equation}
    \begin{split}
        \frac{\partial \LL}{\partial \bv(0)}
         & = \sum_{n=1}^\infty \frac{\partial \LL}{\partial \bv(n)} \frac{\partial \bv(n)}{\partial \bv(0)} + \frac{\partial \LL}{\partial y(0)} \frac{\partial y(0)}{\partial \bv(0)}                                              \\
         & = \sum_{n=1}^\infty \frac{\partial \LL}{\partial \bv(n)} \bA^n + \frac{\partial \LL}{\partial y(0)} \bC^\top                                                                                                             \\
         & = \left(\sum_{n=1}^\infty \left(\bA^\top\right)^n {\frac{\partial \LL}{\partial \bv(n)}^\top} + \bC \frac{\partial \LL}{\partial y(0)}\right)^\top                                                                       \\
         & = \left(\bA^\top \left(\frac{\partial \LL}{\partial \bv(1)}^\top + \sum_{n=2}^\infty \left(\bA^\top\right)^{n-1} {\frac{\partial \LL}{\partial \bv(n)}^\top}\right) + \bC \frac{\partial \LL}{\partial y(0)}\right)^\top \\
         & = \left(\bA^\top {\frac{\partial \LL}{\partial \bz(0)}^\top} + \bC \frac{\partial \LL}{\partial y(0)}\right)^\top                                                                                                        \\
         & = \frac{\partial \LL}{\partial \bz(-1)}.
    \end{split}
\end{equation}
The last three steps are based on Eq.~\eqref{eq:dLdz_full} and Eq.~\eqref{eq:dLdCD}.

\section{Example code}
We provide a minimal Python code below to demonstrate how to implement the recursion $\bv(n+1) = \bA \bv(n) + \bz(n)$ as a differentiable operator in PyTorch.
The forward call is just for reference and not optimised.

\begin{lstlisting}[language=Python, label={lst:ssm}]
import torch
from torch import Tensor
from typing import Tuple
from torch.autograd import Function

class LTIMatrixRecurrence(Function):
    @staticmethod
    def forward(A: Tensor, v0: Tensor, z: Tensor) -> Tensor:
        # A: (M, M)
        # v0: (M,)
        # z(0:N-1): (N, M)
        # returns v(1:N): (N, M)
        # Ideally, implement the forward pass using C++/CUDA extension for efficiency.
        # Here is an inefficient reference implementation in Python.
        v = z.clone()
        v_n = v0
        for n in range(z.shape[0]):
            v[n, :] += A @ v_n
            v_n = v[n, :]
        return v

    @staticmethod
    def setup_context(ctx, inputs, output):
        A, v0, _ = inputs
        v = output
        ctx.save_for_backward(A, v0, v)

    @staticmethod
    def backward(
        ctx, grad_v: Tensor
    ) -> Tuple[Tensor, Tensor, Tensor]:
        A, v0, v = ctx.saved_tensors

        flipped_grad_z = LTIMatrixRecurrence.apply(
            A.T, torch.zeros_like(v0), grad_v.flip(0)
        )

        grad_z = flipped_grad_z.flip(0)
        grad_v0 = A.T @ grad_z[0, :]
        padded_v = torch.cat([v0.unsqueeze(0), v[:-1]], dim=0)
        grad_A = grad_z.T @ padded_v
        return grad_A, grad_v0, grad_z
\end{lstlisting}

\end{document}